# A Quantum Neural Network Computes Entanglement

E.C. Behrman, V. Chandrashekar, Z. Wang, C.K. Belur, J.E. Steck, and S.R. Skinner
Departments of Physics, Computer Science, Electrical and Computer Engineering, and
Aerospace Engineering
Wichita State University, Wichita, KS 67260-0032

**Abstract**

An outstanding problem in quantum computing is the calculation of entanglement, for which no closed-form algorithm exists. Here we solve that problem, and demonstrate the utility of a quantum neural computer, by showing, in simulation, that such a device can be trained to calculate the entanglement of an input state, something neither an algorithmic quantum computer nor a classical neural net can do.

**Introduction**

Recently there has been an explosion of interest in quantum computing. The possibilities seem vast. Beyond the improvements in size and speed is the ability, at least in principle, to do classically impossible calculations. Two aspects of quantum computing make this possible: quantum parallelism and entanglement. While a computational setup can be constructed [1] which makes use only of quantum parallelism (superposition), use and manipulation of entanglement realizes the full power of quantum computing and communication. [2]

The major bottleneck to the use of quantum computers, once they are designed and built, is the paucity of algorithms that can make use of this power. At present there are only two: Shor's factorization [3] and Grover's data base search [4]. It is not yet at all clear that a way will be or can be found to generate efficiently algorithms to solve general problems on quantum computers.

In previous work [1] we have proposed the use of quantum *adaptive* computers to answer this need. An adaptive computer, since it can be trained, adapts to learn and in a sense constructs its own algorithm for the problem from the training set supplied. A *quantum* neural computer should be able to construct algorithms to solve problems which are inherently quantum mechanical – which cannot even be formulated classically - like the calculation of entanglement. We show here that that can be done. Our quantum neural computer uses training methods from the artificial neural network literature to learn to compute a general measure of entanglement. This is a problem which (since there exists, at present, no closed form algorithm) an algorithmic quantum computer cannot solve, and (since the problem is inherently quantum mechanical) no classical computer, neural or not, could solve.

**Quantum neural network**

Two maximally entangled states are the Bell state, $|00\rangle+|11\rangle$, and the EPR state, $|10\rangle+|01\rangle$. Neither can be represented as a product state $(a|1\rangle+b|0\rangle)_A (c|1\rangle+d|0\rangle)_B$. They can be represented by specifying 4 c-numbers, but if we write the state as a product state not all those numbers are independent, since the product is equal to $ac|11\rangle+ad|10\rangle+bc|01\rangle+bd|00\rangle$. So, for example, with the Bell state, one cannot specify complex numbers a,b,c,d such that ac and bd are not zero but ad and bc are. This is also equivalent to saying [5] that the density matrix for entangled states does not trivially factorize.

Thus for an entangled state we must specify the amplitudes in each of the joint states |11>, |10>, |01>, and |00>; *i.e.*, our general state is given by $\alpha_{11}|11>+\alpha_{10}|10>+\alpha_{01}|01>+\alpha_{00}|00>$. (Only three of these are independent, because of normalization.) To encode a mixed state, we write a ket representation with a phase shift, which we then integrate over. So, for example, the mixed state $(|00><00| + |11><11|)/2$ would be represented as $\frac{1}{2}\int_0^{2\pi}\{|00><00| + \exp(i\theta)|11><11|\}d\theta$. These amplitudes then represent the state whose entanglement we wish to measure. These are the inputs to our measure.

Consider a quantum system that evolves in time according to the Hamiltonian:

$$H = K(\sigma_{xA} + \sigma_{xB}) + \sum_i \lambda_i x_i (\sigma_{zA} + \sigma_{zB}) + \frac{m_i \omega_i^2 x_i^2}{2} + \frac{m_i \dot{x}_i^2}{2} \quad (1)$$

i.e., the standard spin-boson Hamiltonian [6], with two two-level systems (TLSs) A and B, and a set of oscillators (bosons){i}. This is a very general form: If the set is made infinite any desired influence, including dissipation, can be modeled [7]. Here we take as our physical model a SQUID system with applied microwave fields [8]. (In fact it turns out that we need only use a single applied field at very low frequency, plus the tunneling amplitude K, for training the net to compute entanglement.) We imagine that the two TLS are far enough apart that they do not interact directly, though this modification would be easy to include. Note that the Hamiltonian under which we evolve the system in time is not the same as that which would have had to have existed in order to create the entanglement in the first place.

We write the time development Green's function operator for the system using the operator calculus developed by Feynman [9]:

$$G[S_A(t_f), S_B(t_f), t_f; S_A(0), S_B(0), 0] = \int D\, S_A(t)\, D\, S_B(t) \prod_i D\, x_i(t)\, \exp\{\frac{i}{\hbar}\int_0^{t_f} dt\, L(S_A, S_B, \{x_i\}, t)\}$$

(2)

where $S_{A,B}(t) = \pm 1$ is the discrete variable labeling the state of each TLS, corresponding to the operators $\sigma_{zA,B}$. L is the Lagrangian of the system. All possible paths for the system are summed over, for each of the variables {S} and {x}, and the time is integrated from 0 to the final time $t_f$. The state of the system at the final time is given by:

$$|\Psi(S_A(t_f), S_B(t_f), t_f)> = G[S_A(t_f), S_B(t_f), t_f; S_A(0), S_B(0), 0]\, |\Psi(S_A(0), S_B(0), 0)>. \quad (3)$$

Each of the four amplitudes for the possible states of the system at the final time, $|\Psi(S_A(t_f) = \pm 1, S_B(t_f) = \pm 1)>$, is a sum over the contributions, weighted by $\alpha_{ij}$, from each of the initial states.

For a traditional artificial neural network, the calculated activation $\phi_i$ of the *i*th neuron is performed on the signals $\{\phi_j\}$ from the other neurons in the network, and is given by $\phi_i = \sum_j w_{ij} f_j(\phi_j)$, where $w_{ij}$ is the weight of the connection from the output of neuron *j* to neuron *i*, and $f_j$ is a bounded differentiable real valued neuron activation function for neuron *j*. Individual neurons are connected together into a network to process information from a set of input neurons to a set of output neurons. The network is an operator $F_W$ on the input vector

$\phi_{input}$. $F_W$ depends on the neuron connectivity weight matrix **W** and propagates the information forward in "space" to calculate an output vector $\phi_{output}$: $\boldsymbol{\varphi}_{output} = \boldsymbol{F_W}\boldsymbol{\varphi}_{input}$. Equation (3) maps the initial state (input) to the final state (output) in much the same way. We now discretize the path integrals in equation (3) to obtain

$$|\Psi(\{S\},\{x\},t_f)\rangle = \lim_{n\to\infty}\int dS_A(0)dS_B(0)\cdots dS_A(t_n)dS_B(t_n)\prod_i dx_i(0)\cdots dx_i(t_n)\exp\left\{\frac{i}{\hbar}\int_0^{t_n} L\right\} \quad (4)$$

If we take n to be finite the n intermediate states can be considered to be the states of n quantum neurons, one at each time slice. The weights have different values at different time slices; thus we can think of the timeline distribution points as virtual neurons, with the neurons distributed in time rather than space. Since we are uninterested in the detailed dynamics of the bosons, and in order to simplify the computation, we integrate out the {x} degrees of freedom, which generates non-local interactions between the neurons at different times. This is the quantum temporal neural net. Figure 1 shows a schematic of the quantum temporal net as it evolves in time. Complete details of the network are given in [1]. The adjustable parameters for the training are the set $\{\lambda_i, m_i, \omega_i\}$ and the tunneling amplitude K, all of which can be adjusted experimentally for the SQUID system under consideration [10].

**Measurement of entanglement**

We take as our entanglement measure the square of the 2-qubit correlation function at the final time, that is, $\langle\sigma_{zA}(t_n)\sigma_{zB}(t_n)\rangle^2$. This is the quantity that we wish our network to train to give the entanglement of the input state. Physically we imagine that our net has been prepared in the input state, which we then evolve in time according to the weighted Hamiltonian described above, after which we measure the correlation function. Many measures of entanglement have been proposed [11-13]. Vedral et al. [12] have proposed a general criterion: that the entanglement of a state, $E(\rho^*)$, specified by its density matrix $\rho^*$, be defined as $E(\rho^*) = \min_{\rho\in D} Dist(\rho^*\|\rho)$, where D is the subset of density matrices in the space which are disentangled. "Dist" is some measure of "distance" between the two density matrices $\rho^*$ and $\rho$ such that $E(\rho^*)$ satisfies the three conditions: first, that $E(\rho^*) = 0$ iff $\rho^*$ is separable; second, that local unitary operations leave $E(\rho^*)$ invariant; and third, that classical correlation does not increase $E(\rho^*)$. The idea is that the entanglement of a state can be defined as its distance from the closest possible state that is not entangled. They propose two different distance measures: the von Neumann relative entropy $Dist_S(\rho^*\|\rho) = tr\left\{\rho^*\ln\frac{\rho^*}{\rho}\right\}$, which is a generalization of the more familiar von Neumann mutual information [11] and to which it is equal for pure states, and the Bures metric $Dist_B(\rho^*\|\rho) = 2 - 2\sqrt{\left[tr\left\{\sqrt{\rho}\rho^*\sqrt{\rho}\right\}^{1/2}\right]^2}$. It is unclear how one is to choose between the two or indeed among the many possible different measures that meet their conditions. (However, once one chooses a measure, a minimum distance can be found, since we have a general formulation for an unentangled state, given above as ac|11⟩+ad|10⟩+bc|01⟩+bd|00⟩, with $|a|^2 + |b|^2 = 1 = |c|^2 + |d|^2$; i.e., a minimization problem of

two independent (complex) variables. In a subsequent publication we enlarge on this, and related matters.)

This problem seems tailor made for neural networks, since neural nets can learn a mapping, given a training set. A set of states is shown in Table 1. Clearly the Bell and EPR states are maximally entangled, and product and mixed states are minimally entangled. We include also classically correlated but unentangled states, and a state with partial entanglement. In the sixth column we list the initial classical correlation of the state, $\langle \sigma_{zA}(0)\sigma_{zB}(0) \rangle$. In the last columns are numbers for the entanglement of each state as calculated by the Vedral method, using the von Neumann and Bures metrics. (Note that the Bures metric does not give us that maximally entangled states, the Bell and the EPR, have entanglement equal to one, but 2-√2; this is the distance to the (unentangled) product state $(|0> + b|1>)_A (|0> + b|1>)_B /(1+|b|^2)$.)

The training is done as follows. A starting set of adjustable environmental parameters $\{\lambda_i, m_i, \omega_i\}$, and K are chosen. Each of the inputs is presented, in turn, to the neural net, and an output value $\langle \sigma_{zA}(t_n)\sigma_{zB}(t_n) \rangle^2$ is calculated, by summing the amplitudes for all possible $2^{2n}$ states. The difference between the desired output for that input, and the output calculated, is the error. Using a standard gradient descent algorithm [14] each of the adjustable parameters is adjusted in the direction that reduces the error for each training pair of (input, output), and the process is repeated. Each repetition is an epoch. Results for the training set are presented in Table 2. The training was done with a discretization of n=3. The RMS error for the set, after 1000 epochs, is essentially zero. The results do not change substantially for larger values of the discretization parameter n, which, even for n=3, is large enough that the discretization error caused by using the short-time form of the propagator is negligible. We then take the final trained values for the various adjustable parameters, and test the net on a new set of inputs, without training (i.e., only a single epoch.) Results for one representative testing set are shown in Table 3. The error for the testing set is also essentially zero. A large number of permutations of possible states have been tried with exactly similar results. The quantum neural net has learned to compute a general measure of entanglement.

The training for the partially entangled system P deserves some further comment. We noted, above, that there is no general agreement on what the entanglement of P is, though it ought to lie somewhere between 0 and 1. Therefore we trained the network for various different target values for the entanglement of P. In Figure 2 we show the output results for this state as a function of the desired value. (Each of these runs used the entire training set, and results for the other training pairs were not significantly different.) We can say that in some sense the value of 0.44 is the "natural" value for the entanglement of this state, at least as computed by this net.

To demonstrate that entanglement is fundamentally a quantum calculation we have also performed the same procedure using a standard neural net program, NeuralWorks [15]. These results are also shown in Tables 2 and 3. Inputs were the normalized amplitudes for each of the four states of the two-qubit system, just as for the quantum neural net; desired outputs were also the same. Adding additional hidden layers, or neurons, does not improve performance: doubling the number produces an identical RMS error. Increasing the size of the training set also does not help, unless the set tested on is essentially equivalent to the set trained on. We tried also a bootstrap train/test, in which we take the entire set of training and testing pairs, then remove each pair in turn, training on the remaining pairs; we then test on the one removed. In each case, the net trained the set of six to essentially zero error, but RMS errors on the tested pair were large: 0.34, 0.62, 0.11, 0.17, 0.48, 0.20, and 0.63. We see that, though a classical neural net can be trained (given a sufficiently large data set), on any given nontrivial test set it does atrociously. That is, it has not learned entanglement as such, but has only fit the data set it was given. In contrast, the quantum neural net did generalize superbly.

**Acknowledgements**

This work was supported by the National Science Foundation, Grant #ECS-9820606. We also acknowledge the support of Kansas NSF Cooperative Agreement EPS-9874732 and the Wichita State University High Performance Computing Center.

| State | Relative amplitudes of | | | | Classical Correlation | Entanglement | |
|---|---|---|---|---|---|---|---|
| | $|00\rangle$ | $|01\rangle$ | $|10\rangle$ | $|11\rangle$ | | von Neumann | Bures |
| Bell | 1 | 0 | 0 | $e^{i\delta}$ | 1 | 1 | 0.59 |
| EPR | 0 | 1 | 1 | 0 | -1 | 1 | 0.59 |
| flat | 1 | 1 | 1 | 1 | 0 | 0 | 0 |
| C | 1 | $\gamma$ | 0 | 0 | $(1-|\gamma|^2)/(1+|\gamma|^2)$ | 0 | 0 |
| P | 0 | 1 | 1 | 1 | -1/3 | 0.32 | 0.27 |
| M | $|00\rangle\langle 00| + |11\rangle\langle 11|$ | | | | 1 | 0 | 0 |

Table1: Some possible states of the two-qubit system. The relative amplitudes (for the ket states) are given without normalization for clarity. The first two are maximally entangled. The second two are product states (flat = $(|0\rangle+|1\rangle)_A(|0\rangle+|1\rangle)_B$ and C=$|0\rangle_A(|0\rangle+|1\rangle)_B$) and thus have zero entanglement; and P is partially entangled. M is a mixed state and cannot be expressed as a ket; its (unnormalized) density matrix is given instead. The classical correlation is computed as $\langle \sigma_{zA}(0)\sigma_{zB}(0) \rangle$. C and M are classically correlated but not entangled.

| State | Quantum neural network | | | NeuralWorks |
|---|---|---|---|---|
| | Initial | Desired | Trained | Trained |
| Bell, $\delta=0$ | 1.0 | 1.0 | 1.0 | 1.0 |
| flat | 0.0 | 0.0 | $1.2 \times 10^{-7}$ | $1.0 \times 10^{-6}$ |
| C, $\gamma=0.5$ | 0.36 | 0.0 | $6.8 \times 10^{-9}$ | $1.0 \times 10^{-7}$ |
| P | 0.11 | 0.44 | 0.44 | 0.44 |

Table 2: Training results for the quantum neural network, on a training set of four, including one completely entangled state, one unentangled state, one classically correlated but unentangled state, and one partially entangled state. Trained values for the tunneling amplitude K are 0.21, 0.25, 0.23, and 0.23 in meV; for the field strengths 0.032, 0.048, 0.035, and 0.00 in meV. Time of evolution was 10 $\hbar$/meV. The last column shows the corresponding values as trained by the standard neural network program NeuralWorks. The RMS error, for both the QNN and NeuralWorks, was essentially zero.

| State | Quantum neural network | | NeuralWorks Output |
|---|---|---|---|
| | Desired | Output | |
| EPR | 1.0 | 1.0 | 0.53 |
| \|00> | 0.0 | $9.2 \times 10^{-10}$ | 0.58 |
| \|10>+0.9\|11> | 0.0 | $4.5 \times 10^{-8}$ | 0.48 |
| $P_2$ | 0.44 | 0.44 | 0.52 |
| M | 0.0 | $9.1 \times 10^{-5}$ | N/A |

Table 3: Representative testing results for the quantum neural network, with NeuralWorks results for comparison. This test set included a maximally entangled state of a type not seen before by the net, the EPR state; a pure but unentangled state; a correlated but unentangled state; a partially entangled state, $P_2$; and a mixed state. $P_2$ is the state $(|00>+|10>+|11>)/\sqrt{3}$. RMS error for the QNN is essentially zero; for NeuralWorks, 0.44. Values for the tunneling amplitude and for the external field for the QNN were as trained above in Table 2.

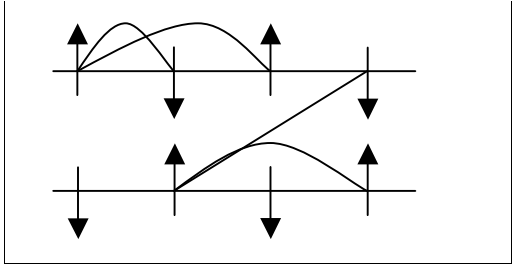

Figure 1: A schematic of the quantum neural network. The two horizontal lines represent the evolution in time of each of the two two-level systems, A and B. Each arrow represents the instantaneous value of a TLS, represented in Equation (4) by $S_{A,B}(t_i)$. Neighboring arrows are connected to each other by tunneling bonds of strength $-\frac{1}{2}\ln[\tan(K\Delta t)]$, from the first term in the Hamiltonian in Equation (1). Each arrow is also connected to each other arrow by a non-local bond, generated by integrating out the boson degrees of freedom; some (but not all) of these are drawn in. Arrows on the far left are the input layer; on the right, the output layer; intermediate arrows are the "hidden layers." Here, the discretization number n=3.

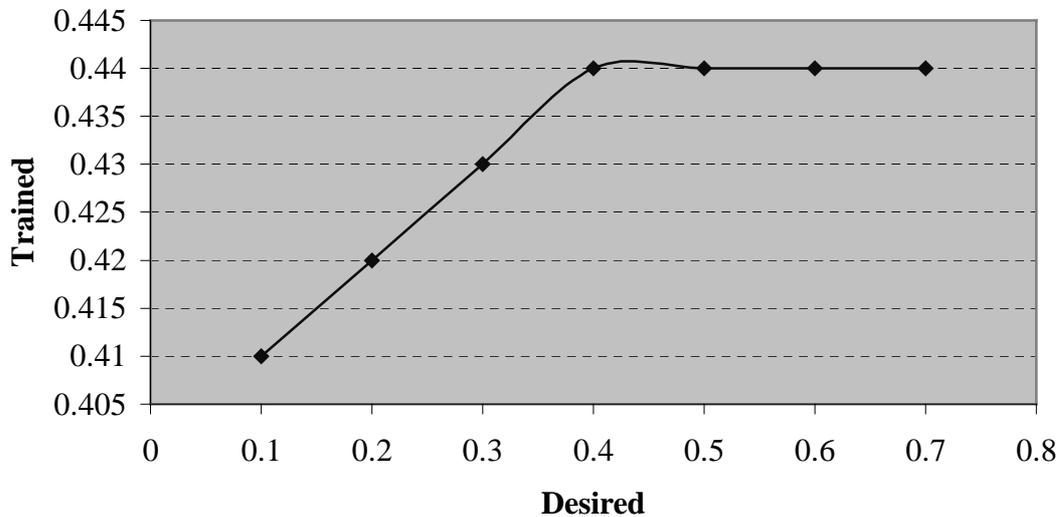

Figure 2: Trained results for the partially entangled state P, as a function of desired value for that state. In each run, represented by a data point on the graph, all four of the training set pairs were trained, and only the desired value for the state P was changed. The line is drawn to guide the eye.